\documentclass[aps,prl,twocolumn,showpacs,superscriptaddress]{revtex4}
\usepackage[english]{babel}
\usepackage{amsmath}
\usepackage{amsfonts}
\usepackage{amssymb}
\usepackage{graphicx}
\usepackage{amsthm}
\usepackage{bm}

\begin{document}

\title{Excitation of $^{229m}$Th at Inelastic Scattering of Low Energy Electrons.}

\author{E.~V.~Tkalya}
\email{tkalya_e@lebedev.ru}

\affiliation{P.N. Lebedev Physical Institute of the Russian
Academy of Sciences, 119991, 53 Leninskiy pr., Moscow, Russia}

\affiliation{National Research Nuclear University MEPhI, 115409,
Kashirskoe shosse 31, Moscow, Russia}

\affiliation{Nuclear Safety Institute of RAS, Bol'shaya Tulskaya
52, Moscow 115191, Russia}

\date{\today}

\begin{abstract}
Excitation of the anomalously low lying nuclear isomer
$^{229m}$Th$(3/2^+, 8.28 \pm 0.17$~eV) in the process of inelastic
electron scattering is studied theoretically in the framework of
the perturbation theory for the quantum electrodynamics. The
calculated cross sections of $^{229m}$Th by the extremely low
energy electrons in the range 9~eV--12~eV for the Th atom and
Th$^{1+,4+}$ ions lie in the range $10^{-25}$--$10^{-26}$ cm$^2$.
Being so large, the cross section opens up new possibilities for
the effective non-resonant excitation of $^{229m}$Th in
experiments with an electron beam or electron (electric) current.
This can be crucial, since the energy of the isomeric state is
currently known with an accuracy insufficient for the resonant
excitation by photons. In addition, the cross section of the time
reversed process is also large, and as a consequence, the
probability of the non-radiative $^{229m}$Th decay via the
conduction electrons in metal is $\approx10^{6}$ s$^{-1}$, that
is, close to the internal conversion probability in the Th atom.
\end{abstract}

\maketitle

Over the past 30 years the anomalously low-lying isomeric level
$3/2^+(E_{\text{is}}<10$~eV) in the $^{229}$Th nucleus has been
the subject of intensive study in experimental and theoretical
ways. Being recently adopted, the value for the energy of the
isomeric state, $E_{\text{is}}=8.28 \pm 0.17$~eV, was obtained in
Ref.~\cite{Seiferle-19}. This result was preceded by a long period
of measurements. During this time the values of $E_{\text{is}}
\leq 100$~eV \cite{Kroger-76}, $1\pm4$~eV \cite{Reich-90},
$3.5\pm1$~eV \cite{Helmer-94}, and $7.8\pm 0.5$~eV \cite{Beck-07}
were obtained. The present value is not the end of the story and a
more refined accuracy is required for the most important
applications of the Thorium isomer, namely, the nuclear clock
\cite{Peik-03,Rellergert-10,Campbell-12,Peik-15} and the nuclear
laser \cite{Tkalya-11,Tkalya-13}.

Studies of the $^{229m}$Th$(3/2^+,8.28 \pm 0.17\,{\text{eV}})$
isomer are also important for many others reasons, including the
fundamental ones. In this paper we refer to the relative effects
of the variation of the fine structure constant and the strong
interaction parameter \cite{Flambaum-06,Litvinova-09,Berengut-09},
control of the isomeric level $\gamma$ decay via the boundary
conditions \cite{Tkalya-18-PRL} or chemical environment
\cite{Tkalya-00-JETPL,Tkalya-00-PRC}, the checking of
exponentiality of the decay law at long times \cite{Dykhne-98},
the detection of the unusual decay of the $^{229}$Th ground state
into the isomeric level in the muonic atom of $^{229}$Th
\cite{Tkalya-16-PRA}, the coherent oscillations between the
components of the hyperfine structure in the Hydrogen-like ion
$^{229}$Th$^{89+}$ \cite{Pachucki-01}, acceleration of the alpha
decay of the $^{229}$Th nucleus via the isomeric state
\cite{Dykhne-96}, and so on.

The present work is focused on the excitation of the $^{229m}$Th
nuclear isomer, which remains one of the problems to be solved.
Excitation of $^{229m}$Th by laser radiation through the electron
shell at the electron bridge process
\cite{Tkalya-92-JETPL,Tkalya-92-SJNP,
Kalman-94,Tkalya-96,Porsev-10-PRL, Muller-19, Borisyuk-19-PRC} is
considered now as the most promising scheme to work with the
$^{229}$Th nucleus. Theoretically, under the resonant conditions
this process provides for the efficient excitation of $^{229}$Th
nuclei. Unfortunately, at present it is extremely difficult to
satisfy these conditions, since not only the nuclear transition
energy, but also the energies and quantum numbers of the excited
states of the Th atom and ions are known with insufficient
accuracy.

In this paper, the excitation of $^{229m}$Th by the low energy
electrons in the Thorium atom and Th$^{1+,4+}$ ions is
investigated through the inelastic electron scattering. The
$^{229}$Th nucleus is screened by the atomic shell consisting of
90 electrons. Therefore, the main scattering channel of low energy
electrons should be caused by the electron shell of the Th atom.
For example, the cross section of electron impact ionization for
the Th atom in the electron energy range of 10--50~eV and the
total elastic scattering cross section in the range 100--200~eV
reach the value of $10^{-15}$~cm$^2$ \cite{Mark-92,Mayol-97}.
Nevertheless, even in this case, the scattered electron can
interact with the nucleus inelastically, since its wave function
near the nucleus has a non-zero amplitude. Note, that the wave
function of the low energy electron, emitted from the valence
shell of the Th atom in the internal conversion process, has
similar properties. (The internal conversion is the main decay
channel of the $^{229m}$Th isomer in the Th atom
\cite{Strizhov-91,Bilous-17,Wense-16,Seiferle-17}.) In the
$^{229m}$Th decay, the excitation energy transfers from the
nucleus to the valence electron because both electron wave
functions --- the bound state and the continuous spectrum state
--- have non-zero amplitudes near the nucleus. The same holds true
for scattering, with the only difference being that both wave
functions belong to a continuum.

The inelastic electron scattering is not resonant. In this point,
it compares favorably with the photon excitation of the $^{229}$Th
nucleus especially at early stages of research, with the exact
value of $E_{\text{is}}$ remaining unknown.

\begin{figure}
 \includegraphics[angle=0,width=0.45\hsize,keepaspectratio]{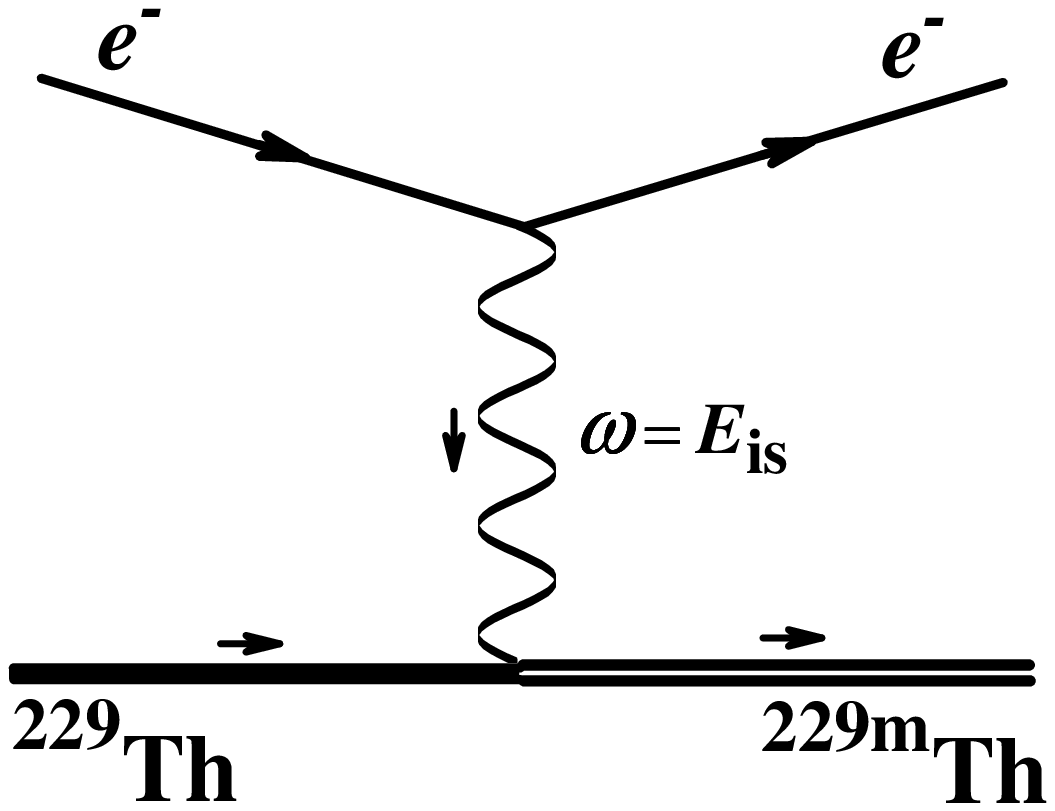}
 \caption{The Feynman diagram of the $^{229}$Th$(e^-,e^-)^{229m}$Th process.}
 \label{fig:FD}
\end{figure}

The cross section of the inelastic electron scattering from a
nucleus is described by the second-order Feynman diagram shown in
Fig.~\ref{fig:FD}
%
%
and in the general case can be written using the Fermi golden
rule,
%
%
\begin{equation}
\sigma=2\pi \int{}d\Omega_{{\bf{p}}_f}
\frac{|\langle{}f|H_{\text{int}}|i\rangle|^2}{v_i}\rho_f,
\label{eq:FermiGold}
\end{equation}
where $v_i=p_i/E_i$ is the speed of the scattering electron,
$p_{i,f}$ and $E_{i,f}=\sqrt{p_{i,f}^2+m_e^2}$ are the momentum
and energy of the electron with mass $m_e$ in the initial, $i$,
and final, $f$, states (the system of units is $\hbar=c=1$),
$\rho_f=p_f E_f /(2\pi)^3$ is the density of final states.

The Hamiltonian of the interacting electron
$j_{fi}^{\varrho}({\bf{r}})$ and nuclear
$J_{fi}^{\varsigma}({\bf{R}})$ currents has the form
%
%
%
\begin{equation}
\langle{}f|H_{\text{int}}|i\rangle = \int{}d^3rd^3R
j_{fi}^{\varrho}({\bf{r}})D_{\varrho\varsigma}(\omega,{\bf{r}}-{\bf{R}})J_{fi}^{\varsigma}({\bf{R}}).
\label{eq:Hint}
\end{equation}
Here the photon propagator in the frequency-coordinate
representation is \cite{Berestetskii-82}
$D_{\varrho\varsigma}(\omega,{\bf{r}}-{\bf{R}})=
-g_{\varrho\varsigma}\exp(i\omega|{\bf{r}}-{\bf{R}}|)/|{\bf{r}}-{\bf{R}}|$,
where $g_{\varrho\varsigma}$ is the metric tensor,
$\omega=E_i-E_f$ is the energy transferred from the electron to
the nucleus. This energy is equal to the isomeric state energy,
i.e. $\omega=E_{\text{is}}$.

Taking into account the Siegert's theorem (see, for example,
\cite{Eisenberg-70}) and the long-wave approximation for the
photon-nucleus interaction, one can simplify the Hamiltonian in
Eq.~(\ref{eq:Hint}) by expanding it in the series over the
electric, $EL$, and magnetic, $ML$, multipoles
%
%
\begin{eqnarray*}
\langle{}f|H_{\text{int}}^{E(M)}|i\rangle
&=&4\pi{}i\omega\sum_{LM} \int{}d^3rd^3R
{\bf{j}}_{fi}({\bf{r}})\cdot{}{\cal{B}}^{E(M)}_{LM}(\omega{}r)\times\nonumber\\
&&\qquad\qquad{\cal{A}}^{E(M)}_{LM}(\omega{}R)\cdot
{\bf{J}}_{fi}({\bf{R}}), \label{eq:HintL}
\end{eqnarray*}
where the dot means the scalar product of the vectors, and
${\cal{A}}^{E(M)}_{LM}$ and ${\cal{B}}^{E(M)}_{LM}$ are the
well-known vector potentials \cite{Eisenberg-70}
%
%
\begin{eqnarray*}
{\cal{A}}^{E}_{LM}(\omega{}R)&=&\sqrt{\frac{L+1}{2L+1}}
{\bf{Y}}_{LL-1,M}({\bf{\Omega}}_{\bf{R}})j_{L-1}(\omega{}R)-\nonumber\\
&&\sqrt{\frac{L}{2L+1}}{\bf{Y}}_{LL+1,M}({\bf{\Omega}}_{\bf{R}})j_{L+1}(\omega{}R),\nonumber\\
{\cal{A}}^{M}_{LM}(\omega{}R)&=&{\bf{Y}}_{LL,M}({\bf{\Omega}}_{\bf{R}})j_{L}(\omega{}R).
\label{eq:AL}
\end{eqnarray*}
The potential ${\cal{B}}^{E(M)}_{LM}$ is obtained from
${\cal{A}}^{E(M)}_{LM}$ by replacing the Bessel spherical function
$j_L(x)$ with the Hankel spherical function of the first kind
$h_L^{(1)}(x)$ \cite{Abramowitz-64},
${\bf{Y}}_{LJ;M}({\bf{\Omega}}_{\bf{R}})$ is the vector spherical
harmonics \cite{Varshalovich-88}.

Using the standard parametrization, the nuclear current is written
as
%
%
\begin{eqnarray*}
\lefteqn{ \left|
\int{}d^3R\,{\bf{J}}_{fi}({\bf{R}}){\cal{A}}^{E(M)}_{LM}(\omega{}R)\right|=}\\
&&{} \frac{\omega^L}{(2L+1)!!}\sqrt{\frac{L+1}{L}}
|\langle{}J_fM_f|\hat{{\cal{M}}}^{E(M)}_{LM}|J_iM_i|\rangle{}|,
\label{eq:ME_J}
\end{eqnarray*}
where the introduced matrix element of the nuclear current
operator $\hat{{\cal{M}}}^{E(M)}_{LM}$ between the states with
spins $J_i$ and $J_f$ is connected with the reduced probability of
the nuclear $E(M)L$ transition by the relation
%
%
\begin{eqnarray*}
\lefteqn{B(E(M)L;J_i\rightarrow{}J_f)=}\\
&&{}\frac{1}{2J_i+1}\sum_{M_i,M_f,M}|\langle{}J_fM_f|\hat{{\cal{M}}}^{E(M)}_{LM}|J_iM_i|\rangle|^2.
\label{eq:ME_J}
\end{eqnarray*}

In the electron current
$$
{\bf{j}}_{fi}({\bf{r}}) = e\psi_{{\bf{p}}_f\mu_f}^{(-)}
{\bm{\alpha}} \psi_{{\bf{p}}_i\mu_i}^{(+)},
$$
$e$ is the electron charge, ${\bm{\alpha}}=\gamma^0{\bm{\gamma}}$,
and $\gamma^{i}$ $\{i=0,1,2,3\}$ are the Dirac matrices,
$\psi_{{\bf{p}}_i\mu_i}^{(+)}$ denotes the wave function of the
initial state of the electron with the momentum ${\bf{p}}_i$ and
the projection of the electron spin $\mu_i$ on the direction
${\bm{\nu}}_i={\bf{p}}_i/p_i$, and $\psi_{{\bf{p}}_f\mu_f}^{(-)}$
is wave function of the final state with the momentum ${\bf{p}}_f$
and the projection $\mu_f$ of the spin on the direction
${\bm{\nu}}_f={\bf{p}}_f/p_f$. These wave functions are the exact
solutions of the Dirac equations which asymptotically go over into
a superposition of a plane wave with a diverging and converging
spherical wave \cite{Berestetskii-82}. The explicit form of
$\psi_{{\bf{p}}\mu}^{(\pm)}$ is as follows
%
%
\begin{equation}
\psi_{{\bf{p}}\mu}^{(\pm)} =4\pi \sum_{j,l,m}
\psi_{E,j,l,m}(x)\left(\Omega^*_{jlm}({\bm{\nu}})\upsilon^{\mu}({\bm{\nu}})\right)e^{(\pm)i\delta_{lj}},
\label{eq:WF}
\end{equation}
where $x=r/a_B$, $a_B$ is the Bohr radius, $j$ and $l$ are the
total and orbital angular momenta of the electron, $m$ is the
projection of $j$ onto the quantization axis. The functions
$\psi_{E,j,l,m}(x)$ in Eq.~(\ref{eq:WF}) are
%
%
\begin{eqnarray*}
\psi_{E,j,l,m}(x) =\frac{1}{pa_B}\sqrt{\frac{E+m_e}{2E}} \left(
\begin{array}{r}
g_{lj}(x)\Omega_{jlm}({\bf{r}}) \\
-if_{l'j}(x)\Omega_{jl'm}({\bf{r}})
\end{array} \right),
\label{eq:WF_G-F}
\end{eqnarray*}
and $l'=2j-l$. The large $g_{lj}$ and small $f_{l'j}$ components
of $\psi_{E,j,l,m}(x)$ are the numerical solutions of the Dirac
equations for the electron energies $E>m_e$
%
%
\begin{equation}
\left.
\begin{array}{ll}
g'(x)+\cfrac{1+\kappa}{x}g(x)-
\cfrac{1}{e^2}\left(\cfrac{E}{m_e}+1-\cfrac{V(x)}{m_e}\right)f(x) =0,\\
f'(x)+\cfrac{1-\kappa}{x}f(x)+
\cfrac{1}{e^2}\left(\cfrac{E}{m_e}-1-\cfrac{V(x)}{m_e}\right)g(x)
=0.
\end{array}
\right. \label{eq:EqDirac}
\end{equation}
The functions $g_{i,f}(x)$ in Eq.~(\ref{eq:EqDirac}) are
normalized at $x\rightarrow\infty$ with the condition
$g(x)=\sin(pa_Bx+\varphi_{lj})/x$, where $\varphi_{lj}$ is a
phase, $\kappa=l(l+1)-j(j+1)-1/4$.

The potential energy $V(x)$ of the electron in
Eq.~\ref{eq:EqDirac} is the sum of the electron shell potential
energy $V_{\text{shell}}(x)$ and that of the unscreened nucleus
$V_{\text{nucl}}(x)$. Under the standard assumption that a nucleus
with the atomic number $A$ and the charge $Z$ is represented by a
uniformly charged sphere of the radius $x_{R_0} = R_0/a_B$, where
$R_0 = 1.2A^{1/3}$ fm we conclude that the electron potential
energy in its potential is
$V_{\text{nucl}}(x) =
-{\cal{E}}_0 (Z/2x_{R_0}) [3-(x/x_{R_0})^2]$
for $0\leq{}x\leq{}x_{R_0}$, and
$V_{\text{nucl}}(x) =-{\cal{E}}_0 Z/x$
for $x\geq{}x_{R_0}$ where ${\cal{E}}_0=m_ee^4$ is the atomic unit
of energy.

\begin{figure}
 \includegraphics[angle=0,width=0.98\hsize,keepaspectratio]{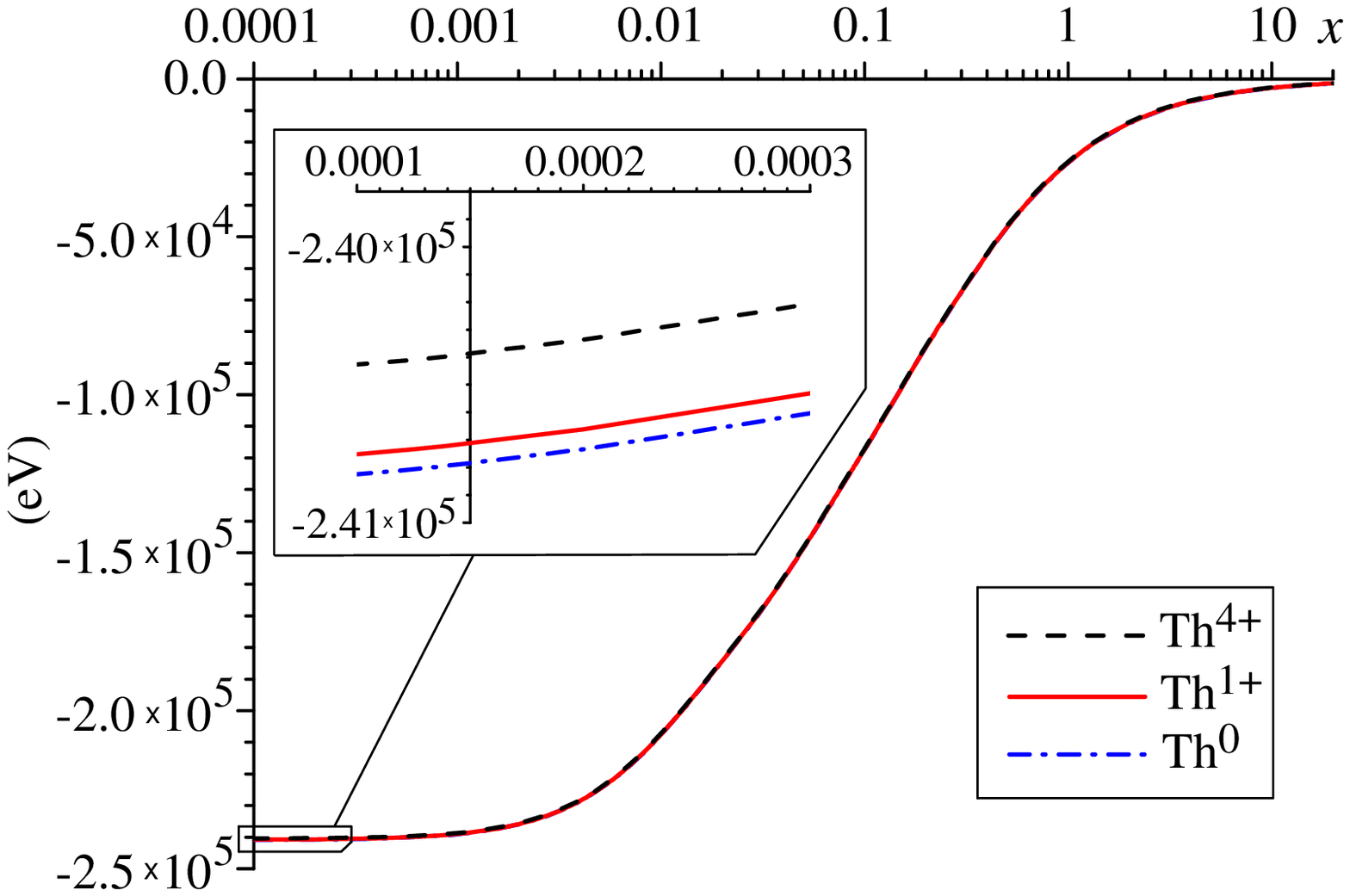}
 \caption{The electron shell potential for the Th atom and the Th$^{1+,4+}$ ions.
 }
 \label{fig:ElectronShellPotential}
\end{figure}

The electron shell potential has been found as follows. At the
first stage, the electron density $\rho_e(x)$ in the Thorium atom
(Th$^0$) and Th$^{1+,4+}$ ions (see an example in
Ref.~\cite{Tkalya-19-PRC-IC_Rydb}) was calculated within the DFT
theory \cite{Nikolaev-15,Nikolaev-16} through the self-consistent
procedure. At the second stage, the radial component of the
electric field as a function of $x$ was found by the numerical
integration of $\rho_e(x)$. And at the third stage, the electron
shell potential was obtained by the numerical integration of the
electric field. The resulting electron shell potentials for
Th$^{0,1+,4+}$ are shown in Fig.~\ref{fig:ElectronShellPotential}.
%
%
The potential energy $V_{\text{shell}}(x)$ for Th$^{0,1+,4+}$ is
$e$ times the corresponding functions in
Fig.~\ref{fig:ElectronShellPotential}.

Substituting the expressions obtained for the currents in
Eq.~\ref{eq:FermiGold}, averaging over the initial states, and
summing over all final states, we arrive at the following
scattering cross section,
%
%
\begin{eqnarray}
\sigma&=&(4\pi{}e)^2 a_B^2
\frac{p_f}{p_i}\frac{E_i+m}{p_i^2} \frac{E_f+m}{p_f^2}\times\nonumber \\
&&\sum_{L} \sum_{{\cal{T}}=E,M} \sum_{l_i,j_i \atop{}l_f,j_f}
\frac{\omega^{2L+2}}{[(2L+1)!!]^2}(2j_i+1) \times \nonumber \\
&&\left(C^{j_f1/2}_{j_i1/2L0}\right)^2
B({\cal{T}}L,J_i\rightarrow{}J_f)
|{\text{\textsl{m}}}_{f,i}^{{\cal{T}}L}|^2 ,
\label{eq:CS}
\end{eqnarray}
where $C^{j_f1/2}_{j_i1/2L0}$ is the Clebsch-Gordan coefficient,
${\text{\textsl{m}}}_{f,i}^{{\cal{T}}L}$ stands for the electron
matrix elements
%
%
\begin{eqnarray}
{\text{\textsl{m}}}_{fi}^{EL} &=& \int_0^{\infty}
h_L^{(1)}(\omega{}a_Bx)[g_i(x)g_f(x)+ f_i(x)f_f(x)]x^2dx, \nonumber\\
{\text{\textsl{m}}}_{fi}^{ML} &=& \cfrac{\kappa_i+\kappa_f}{L}
\int_0^{\infty}h_L^{(1)}(\omega{}a_Bx)[g_i(x)f_f(x)+ \nonumber\\
&&\qquad\qquad\qquad\quad{}f_i(x)g_f(x)]x^2dx.
\label{eq:EME}
\end{eqnarray}
The summation over the orbital momenta $l_i$ and $l_f$ is
performed in Eq.~(\ref{eq:CS}), since the calculation should take
into account all possible combinations of angular momenta and the
parity selection rules. It can be done by means of the well-known
representation for the Clebsch-Gordan coefficient through the $6j$
symbol \cite{Bohr-98-I}
\begin{equation*}
\left(C^{j_f 1/2}_{j_i 1/2 L 0}\right)^2 =(2l_i+1)(2j_f+1)
\left(C^{l_f 0}_{l_i 0 L 0} \right)^2 \left\{
\begin{array}{ccc}
 l_i & 1/2 & j_i \\
 j_f &  L  & l_f
\end{array}
\right\}^2. \label{eq:CGC}
\end{equation*}

We are concerned with the energy region where the kinetic energy
of the electron in the initial state $E_e=E_i-m_e$ satisfies the
condition $E_e \ll m_e$. In that case, the cross section
(\ref{eq:CS}) is simplified, taking the form
%
%
\begin{eqnarray}
\sigma_{E(M)L}&=&4e^2\lambda^2_{\gamma_{\text{is}}}
\left(\frac{E_e}{E_{\text{is}}}\right)^{-3/2}
\left(\frac{E_e}{E_{\text{is}}}-1\right)^{-1/2} \times \nonumber \\
&&\frac{B(E(M)L;J_i\rightarrow{}J_f)}{a_B^{2L}}\times \nonumber \\
&&{}\sum_{l_i,j_i \atop{}l_f,j_f}
\frac{(2li+1)(2j_i+1)(2j_f+1)}{(2L+1)^2} \times \nonumber \\
&&\left(C^{l_f0}_{l_i0L0}\right)^2 \left\{
\begin{array}{ccc}
 l_i & L & l_f \\
 j_f &1/2& j_i
\end{array}
\right\}^2
\left|\tilde{{\text{\textsl{m}}}}^{E(M)L}_{fi}\right|^2,
\label{eq:CS_NR}
\end{eqnarray}
where $\lambda_{\gamma_{\text{is}}}=2\pi/E_{\text{is}}$ is the
wavelength of the isomeric nuclear $\gamma$ transition. The
electron matrix elements in Eq.~(\ref{eq:CS_NR}) are
%
%
\begin{equation}
\begin{split}
\tilde{{\text{\textsl{m}}}}_{fi}^{EL} &= \int_0^{\infty}
[g_i(x)g_f(x)+ f_i(x)f_f(x)]\frac{dx}{x^{L-1}},\\
\tilde{{\text{\textsl{m}}}}_{fi}^{ML} &=
\cfrac{\kappa_i+\kappa_f}{L}
\int_0^{\infty}[g_i(x)f_f(x)+f_i(x)g_f(x)]\frac{dx}{x^{L-1}}.
\end{split}
\label{eq:EME_NR}
\end{equation}
In the case of the $ML$ transition, one needs to change
$l_i\rightarrow{}l'_i=2j_i-l_i$ in formulas
(\ref{eq:CS_NR})--(\ref{eq:EME_NR}).

Two remarks are in order about the numerical solution of
Eq.~(\ref{eq:EqDirac}) and the calculation of the integrals in
Eq.~(\ref{eq:EME}). First, the numerical solution of the Dirac
Eq.~(\ref{eq:EqDirac}) for the wave function of the final state
reaches the asymptotic behavior very slowly near the reaction
threshold at $E_e=8.3$--8.5~eV. Here, the energy of the scattered
electron is very small, and the wavelength, respectively, is
large. Therefore, it is necessary to carry out calculations up to
$x\approx 2\times10^4$ for the correct normalization of the wave
function. Second, the integrals in the matrix elements formally
diverge in Eq.~(\ref{eq:EME}), and converge in
Eq.~(\ref{eq:EME_NR}). In our case, when $\omega{}a_B\approx
1/450$, the approximation
$h_L^{(1)}(\omega{}a_Bx)\approx{}-i(2L-1)!!/(\omega{}a_Bx)^{L+1}$
\cite{Abramowitz-64} used to derive Eq.~(\ref{eq:EME_NR}) is valid
in the region $0 \leq{}x\lesssim 50$. Nevertheless, the use of
formulas (\ref{eq:EME_NR}) is quite correct. Integration over
large values of the variable, $x\geq50$, does not lead to a
divergence of the integrals in Eq.~(\ref{eq:EME}) due to the fast
oscillations of the integrand. The characteristic ``period'' of
these oscillations $x_{\lambda_e}=\lambda_e/a_B \leq8$ is
determined by the wavelength of the electron in the initial state.
In the energy range $E_e\geq{} E_{\text{is}}$, the condition
$\lambda_e\leq 8a_B$ is always satisfied, leading to the fast
convergence of the integrals. In addition, it follows from the
numerical calculation (see the Supplement) that the largest
contribution (95\%) to the electron matrix elements (\ref{eq:EME})
and (\ref{eq:EME_NR}) is due to the integration from a much
smaller area, $0\leq{}x\lesssim 0.01$.

Currently, there are no experimental data on the excitation of the
low-lying nuclear states by the low energy electrons. That is why
the model was tested using the atomic data. The excitation cross
section of the Th atom in the $7s_{1/2}\rightarrow7p_{1/2}$
transition calculated by formula (\ref{eq:CS_NR}) is in good
agreement with the similar data for the Pb atom \cite{Eck-20}  if
one replaces in Eq.~(\ref{eq:CS_NR}) a nuclear matrix element with
the indicated atomic $E1$ matrix element.

\begin{figure}
 \includegraphics[angle=0,width=0.98\hsize,keepaspectratio]{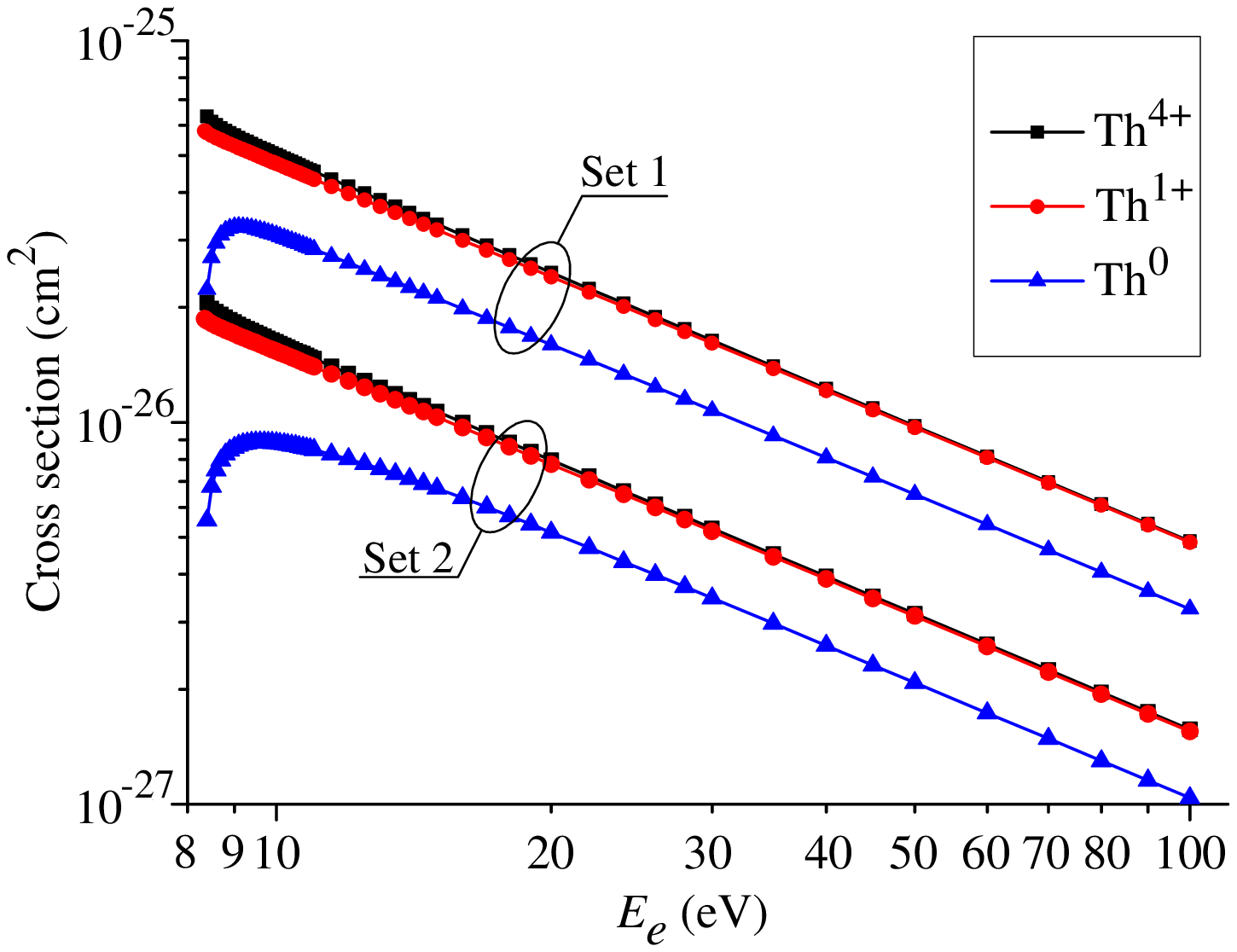}
 \caption{The electron inelastic scattering cross section of the process
 $^{229}$Th$^{0,1+,4+}(e^-,e^-)^{229m}$Th$^{0,1+,4+}$
 with two data sets: Set 1 --- $B_{\text{W.u.}}(M1,5/2^+\rightarrow 3/2^+)
 = 0.031$ and  $B_{\text{W.u.}}(E2,5/2^+\rightarrow 3/2^+)=11.7$,
 Set 2 --- $B_{\text{W.u.}}(M1,5/2^+\rightarrow 3/2^+)=0.0076$,
 $B_{\text{W.u.}}(E2,5/2^+\rightarrow 3/2^+)=27$.}
 \label{fig:CS}
\end{figure}

The cross sections for two sets of the reduced probabilities
$B_{\text{W.u.}}(M1;J_i\rightarrow{}J_f)$ and
$B_{\text{W.u.}}(E2;J_i\rightarrow{}J_f)$ are shown in
Fig.~\ref{fig:CS}. The first set is based on the experimental data
\cite{Bemis-88,Gulda-02,Barci-03,Ruchowska-06} for the $M1$ and
$E2$ transitions between the rotation bands $3/2^+[631]$ and
$5/2^+[633]$ in the $^{229}$Th nucleus found with Alaga rules in
Refs.~\cite{Dykhne-98,Tkalya-15-PRC}). The second set utilizes
$B_{\text{W.u.}}$ for the $M1$ and $E2$ transitions from
Ref.~\cite{Minkov-17}, resulting from the computer calculation
made in the compliance with the modern nuclear models.
%
%

Within the considered region of small $E_e$, the radial electron
wave functions are proportional to $E_e^{1/4}$ in the initial
state, and to $(E_e-E_{\text{is}})^{1/4}$ in the final state in
the Coulomb potential \cite{Landau-QM}. As a result, for the
Th$^{1+,4+}$ cross section, we obtain the dependence
$\sigma\propto{}1/E_e$, Fig.~\ref{fig:CS}.

The difference in the shape of the cross sections for the atom and
the ions near the reaction threshold of 8.28~eV is due to the fact
that in case of atom the electron interacts with the nucleus at
relatively short distances ($\lesssim{}a_B$) when it penetrates
into the electron shell, whereas in the case of ions the Coulomb
interaction occurs at larger distances. For the Th atom near the
reaction threshold $\sigma\propto\sqrt{E_e-E_{\text{is}}}$
\cite{Landau-QM} and tends to zero at
$E_e\rightarrow{}E_{\text{is}}$. In the case of the
$^{229}$Th$^{1+,4+}$ ions, the cross section near the reaction
threshold approaches a constant, $\sigma\rightarrow{\text{const}}$
\cite{Landau-QM}.

As for the contributions to the cross section of partial waves,
then, as one would expect \cite{Landau-QM}, the $S$ wave, namely
the $S^{(i)}_{1/2}\rightarrow{}S^{(f)}_{1/2}$ transition, makes
the largest contribution to the $M1$ scattering (see the
Supplement), whereas the $P^{(i)}_{1/2}\rightarrow{}P^{(f)}_{3/2}$
and $P^{(i)}_{3/2}\rightarrow{}P^{(f)}_{1/2}$ transitions are
dominant for the $E2$ component of the cross section (see the
Supplement).

Being very large, the cross section opens up completely new
possibilities for the excitation and study of the low-lying
nuclear isomer $^{229m}$Th. The first possibility is the
excitation in the dense laser plasma with the electron temperature
$T\approx E_{\text{is}}$. The ratio of the number of excited
nuclei ($N_{\text{is}}$) to the number of nuclei in the ground
state ($N_{\text{gr}}$) in a plasma bunch with the electron energy
distribution $f(E_e)$ can be estimated as in \cite{Tkalya-04}
$$
N_{\text{is}}/N_{\text{gr}} \approx{}n_e\tau
\int^{\infty}_{E_{\text{is}}} f(E_e)v(E_e)\sigma(E_e) dE_e.
$$
For the electrons with the Maxwell energy distribution $f(E_e)$,
the reaction rate $\langle{}\sigma(E_e)v(E_e)\rangle$ reaches
$10^{-18}$--$10^{-17}$~cm$^{3}$s$^{-1}$. In the plasma produced by
the laser pulse of the duration $\tau\approx 10^{-8}$--$10^{-9}$~s
on a solid target, the electron density is $n_e\approx
10^{19}$--$10^{20}$~cm$^{-3}$. As a result, we obtain
$N_{\text{is}}/N_{\text{gr}} \approx{}10^{-6}$, which corresponds
to the efficiency of the resonant process of nuclear excitation by
electron capture (NEEC). Indeed, in the NEEC process (see for
example \cite{Goldanskii-76,Doolen-78-PRL,Doolen-78-PRC}), the
nucleus is excited resonantly by the plasma electrons with the
energy of $E_{e_{\text{res}}} \approx{} E_{\text{is}}-|E_b|$,
where $E_b$ is the electron binding energy for the ionized shell
(for simplicity, in the Thorium atom, we consider only the shell
that makes the main contribution to the probability of the
internal conversion, $\Gamma_{\text{is}}^{\text{IC}}$, during the
decay of $^{229m}$Th). For NEEC, the working region of the
electron spectrum equals approximately to the internal conversion
width of the nuclear state, i.e. $\Gamma_{\text{is}}^{\text{IC}}$.
The energy of resonant electrons is $E_{e_{\text{res}}}\approx
2$~eV, because the characteristic binding energy of the valence
electrons in Th is about 6 eV. As a consequence, the effective
NEEC cross section from the electrons with the resonant wave
length
$\lambda_{e_{\text{res}}}=2\pi/\sqrt{2m_eE_{e_{\text{res}}}}$ is
$$
\sigma_{\text{NEEC}} \approx
(\lambda_{e_{\text{res}}}/2)^2\Gamma_{\text{is}}^{\text{IC}}/T
\approx 10^{-25}~{\text{cm}}^2
$$
for the radiation width of the isomeric
transition $\Gamma_{\text{is}}^{\text{rad}}\approx
3.6\times10^{-19}$~eV and the internal conversion coefficient
$\alpha = 1.6\times 10^{9}$. Taking into account the factor
$f(E_{e_{\text{res}}})$ one can obtain the same value
$N_{\text{is}}/N_{\text{gr}} \approx{}10^{-6}$ for the fraction of
the excited nuclei. (That is only to be expected, as, according to
the perturbation theory for the quantum electrodynamics, the
inelastic electron scattering and NEEC are the second-order
processes described by the same Feynman diagram.) For further
studies, Th ions with the $^{229}$Th excited nuclei can be
extracted by an external electric field from a plasma and
implanted into thin film with a wide-gap dielectric material
(SiO$_2$) (see details in
Refs.~\cite{Fominskii-04,Borisyuk-18,Borisyuk-18-arXiv,Lebedinskii-20}).

The second possibility is to excite $^{229m}$Th by the electron
(electric) current in solids or in experiments with the
high-current electron beam \cite{Borisyuk-16}. For the density of
implanted nuclei $\rho_{\text{Th}} = 10^{18}$--$10^{19}$
cm$^{-3}$, the target thickness $h=10$ nm and the current $j_e= 1$
$A$, the rate of electron excitation of the isomeric nuclei in
solids can be estimated as
$$
dN_{\text{is}}/dt \approx{} \rho_{\text{Th}} h j_e \approx
10^5{\text{--}}10^6~{\text{s}}^{-1}.
$$
In the electron beam experiment, the generation of an avalanche of
the secondary electrons with the energies $E_e>E_{\text{is}}$
increases the efficiency of excitation of the $^{229}$Th nuclei.
Note that in these experiments it is not necessary to know the
energy of the nuclear isomeric level $E_{\text{is}}$ with high
accuracy to tune the electron energy, since the excitation process
is non-resonant.

Another interesting opportunity to observe the process discussed
above is to expose the electron shell to the intense laser field
after tunnel ionization \cite{Andreev-19}. In this case the
electrons accelerated back to the ionized Th atom by the laser
field should excite the $^{229}$Th nuclei through the inelastic
scattering and NEEC.

Using the principle of detailed balance \cite{Landau-QM}
$$
(2J_{\text{gr}}+1) p_i^2
\sigma_{{\text{gr}}\rightarrow{\text{is}}} = (2J_{\text{is}}+1)
p_f^2 \sigma_{{\text{is}}\rightarrow{\text{gr}}}
$$
one can also calculate the cross section for the time reversed
process, which is the decay of the isomeric state through the
electron states in the continuum. Such a situation arises, for
example, when the isomer is implanted into metal
\cite{Tkalya-99-JETPL}. In the free electron approximation
\cite{Aschroft-76} the decay probability through the conduction
electrons in metal is given by
$$
W_{(e,e')} \approx{}\rho_{e} v_F \sigma_{fi}.
$$
Here $\rho_{e}$ is the density of conduction electrons, $v_F$ is
the Fermi velocity, and $E_F$ is the Fermi energy. For the
``standard'' metal \cite{Pippard-60} with $\rho_{e} =
6\times10^{22}$ cm$^{-3}$ and $E_F=5.5$~eV we have
$W_{(e,e')}\approx 10^6$~s$^{-1}$ and the half-life of the
$^{229m}$Th isomer is about $10^{-6}$~s. This is comparable to the
half-life of $^{229m}$Th decay via the internal conversion channel
\cite{Strizhov-91,Seiferle-17}.

In conclusion, the excitation cross section of the low lying
isomer in the $^{229}$Th nucleus has been calculated for the
inelastic scattering of the extremely low energy electrons.
Firstly, the cross section turned out to be so large that the
decay probability of $^{229m}$Th in the process of the conversion
on the conduction electrons in metal is close to the probability
of the internal conversion in the Th atom. Secondly, the
calculated cross section provides for the effective excitation of
$^{229m}$Th: a) in the dense laser plasma with the temperature
$T\approx{}E_{\text{is}}$, b) in solids by the electron (electric)
current, c) at the high-current electron beam. Thirdly, this
approach is fundamentally different from the well-known photon
excitation, since it is non-resonant in nature and does not
require that the energy of the excited level should be known. This
is especially valuable at present, when the energy of the isomeric
level is known with the accuracy of several tenths of the electron
volt.

This research was supported by a grant of the Russian Science
Foundation (Project No 19-72-30014).

\section{Supplement}
\subsection{Electron matrix elements}

%
\begin{figure}[h]
 \includegraphics[angle=0,width=0.8\hsize,keepaspectratio]{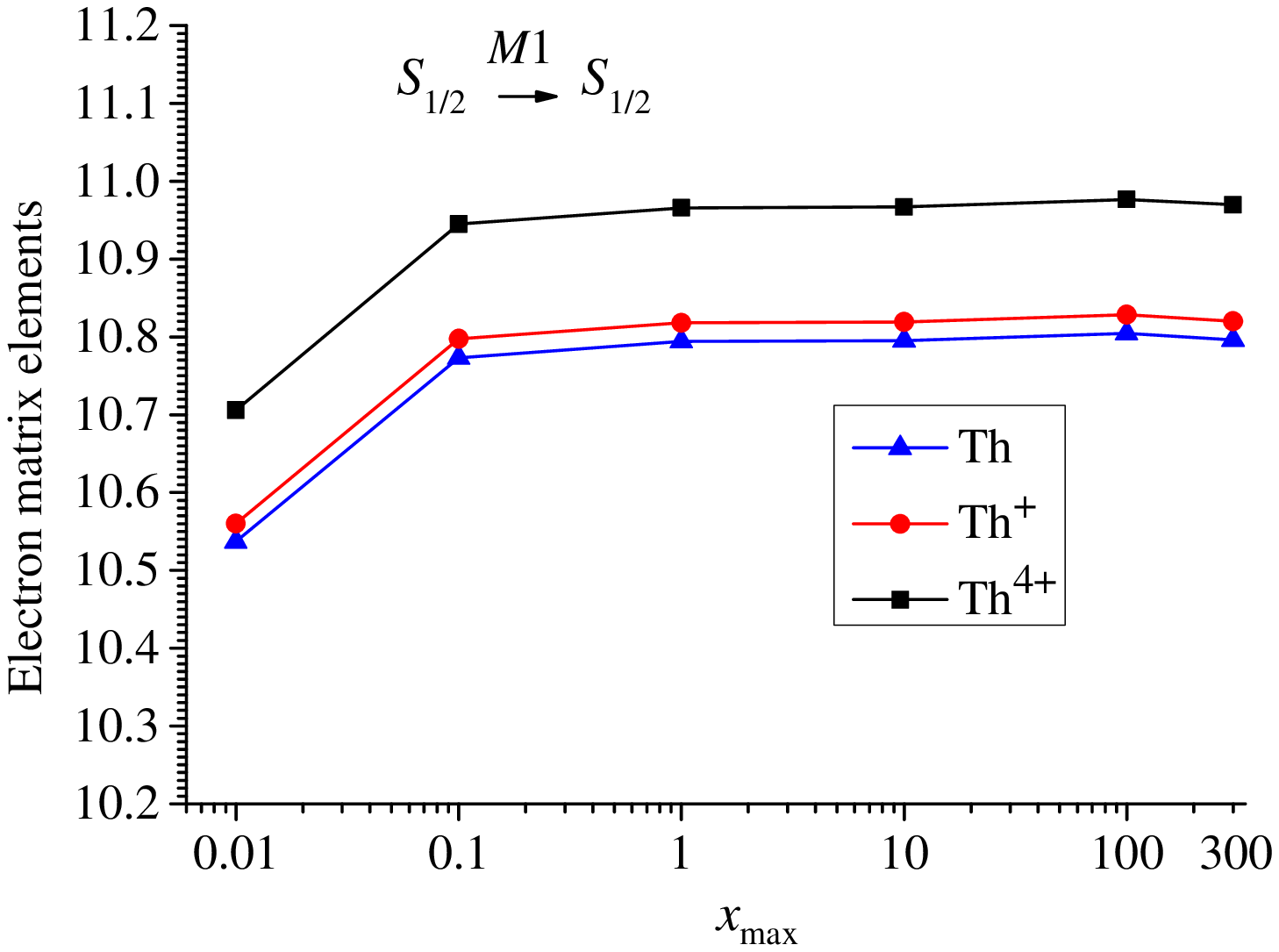}
 \caption{Electron matrix elements as a function of the end
 value, $x_{\text{max}}$, of the integral in Eq.~(\ref{eq:EME_r}).}
 \label{fig:EME}
\end{figure}
As an example, the $M1$ electron matrix elements
%
%
\begin{eqnarray}
\frac{(\omega_Na_B)^2}{3}&\times &(\kappa_i+\kappa_f)
\int_0^{x_{\text{max}}}h_1^{(1)}(\omega{}a_Bx)
[g_i(x)f_f(x)+\nonumber\\
&&f_i(x)g_f(x)]x^2dx.
\label{eq:EME_r}
\end{eqnarray}
for the partial $S_{1/2}\stackrel{M1}\longrightarrow{}S_{1/2}$
transition at the energy of $E_e=20$~eV are shown in
Fig.~\ref{fig:EME}.

Note that the matrix elements in Eq.~(\ref{eq:EME}) multiplied by
the factor $(\omega_Na_B)^{L+1}/(2L-1)!!$ coincide with the
corresponding matrix elements in Eq.~(\ref{eq:EME_NR}) with the
relative accuracy of $10^{-7}$.

It can be seen that the matrix elements in Fig.~\ref{fig:EME}
converge quickly and the main contribution to the integral comes
from the region near the nucleus $0\leq{}x\lesssim0.01$.

\subsection{Partial waves in cross section}

\begin{figure}[h]
 \includegraphics[angle=0,width=0.8\hsize,keepaspectratio]{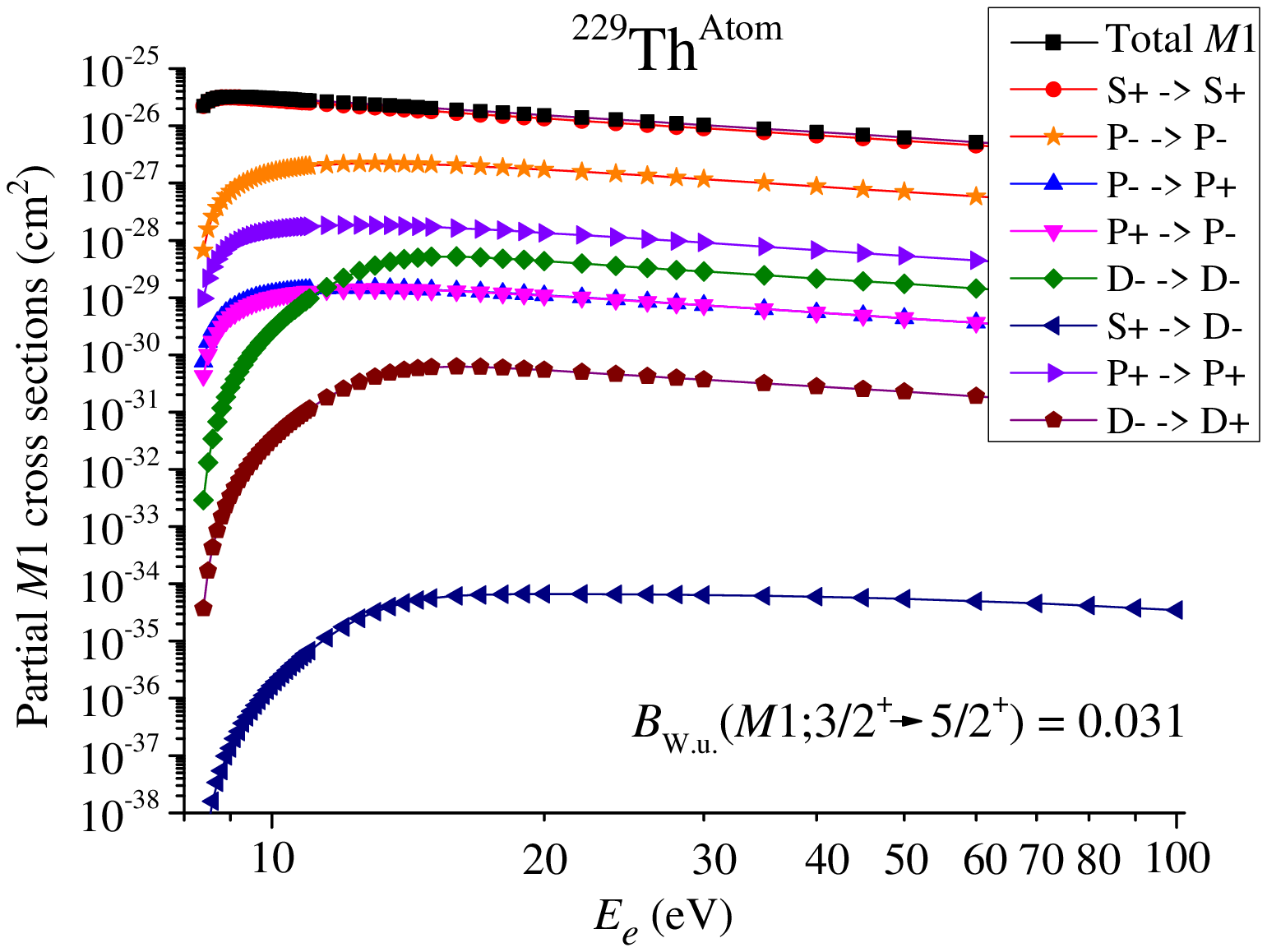}
 \caption{Partial $M1$ cross sections for the $^{229}$Th(e,e')$^{229m}$Th reaction.}
 \label{fig:PCS_Th_M1}
\end{figure}

\begin{figure}[h]
 \includegraphics[angle=0,width=0.8\hsize,keepaspectratio]{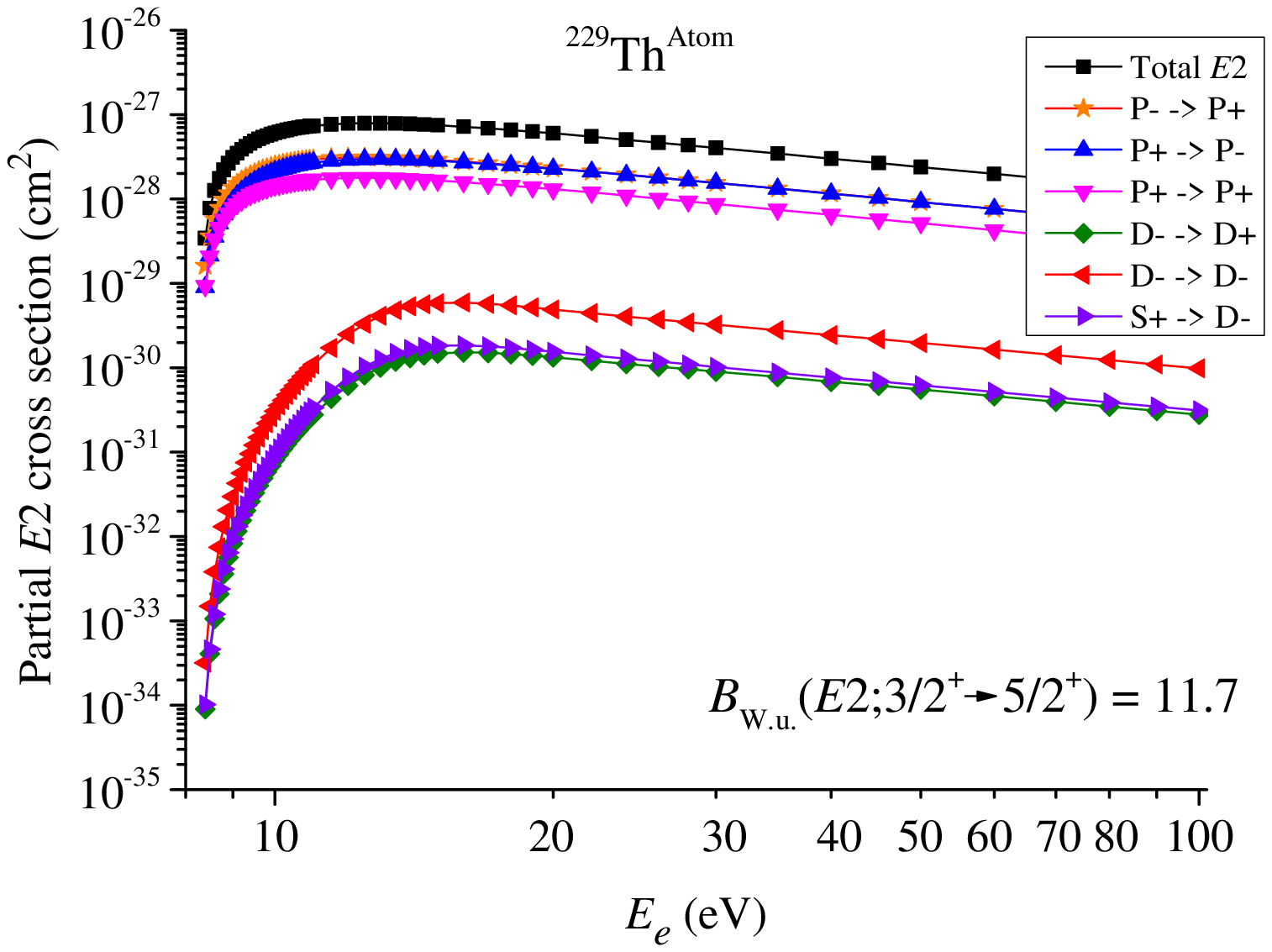}
 \caption{Partial $E2$ cross sections for the $^{229}$Th(e,e')$^{229m}$Th reaction.}
 \label{fig:PCS_Th_E2}
\end{figure}

\begin{figure}[h]
 \includegraphics[angle=0,width=0.8\hsize,keepaspectratio]{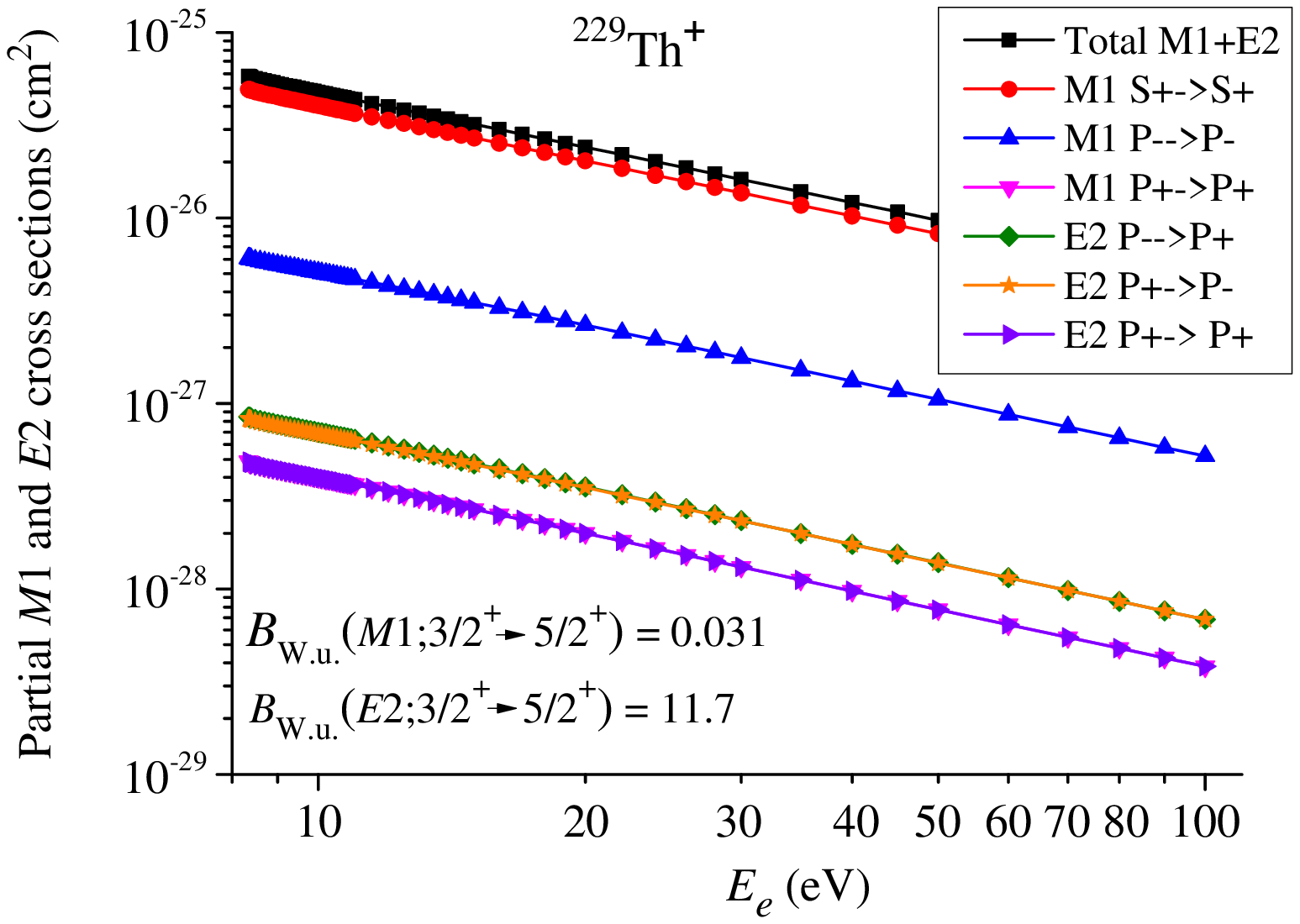}
 \caption{Partial $M1$ and $E2$ cross sections for the $^{229}$Th$^+$(e,e')$^{229m}$Th$^+$ reaction.}
 \label{fig:PCS_Th+_M1-E2}
\end{figure}

\begin{figure}[h]
 \includegraphics[angle=0,width=0.8\hsize,keepaspectratio]{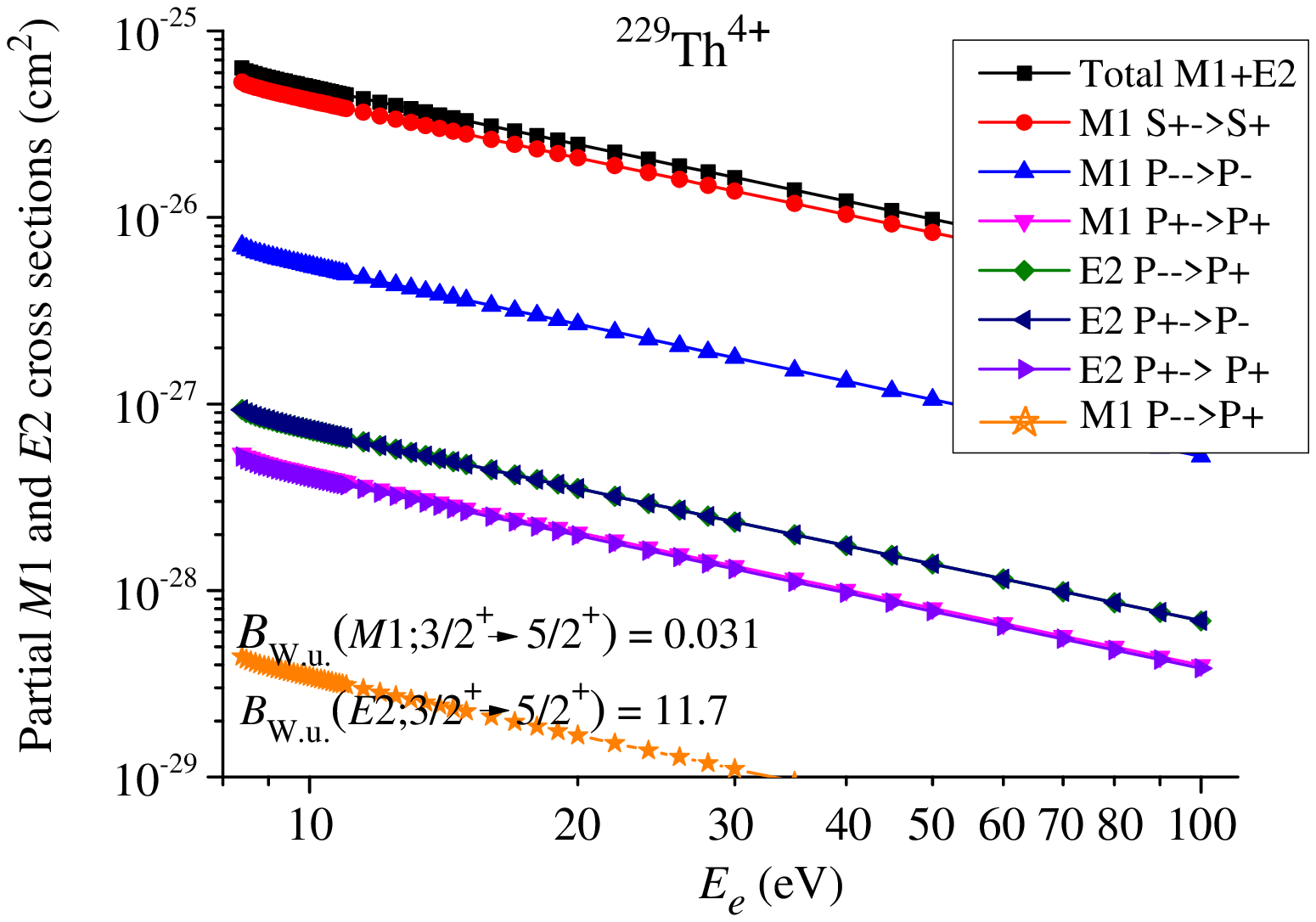}
 \caption{Partial $M1$ and $E2$ cross sections for the $^{229}$Th$^{4+}$(e,e')$^{229m}$Th$^{4+}$ reaction.}
 \label{fig:PCS_Th4+_M1-E2}
\end{figure}

The contributions of various partial waves to the $M1$ and $E2$
cross sections for the Th atom and Th$^+$ and Th$^{4+}$ ions are
presented in Figures
\ref{fig:PCS_Th_M1}--\ref{fig:PCS_Th4+_M1-E2}.
%
%
%
%
%
%
%
%
%

\end{document}